\documentclass[12pt]{article}
\renewcommand{\d}{{\rm d}}
\newcommand{\bl}{\mbox{\boldmath$l$}}

\newcommand{\pmtau}{\,^{\pm}\!\tau}
\newcommand{\pmpi}{\,^{\pm}\!\pi}
\newcommand{\pmbpi}{\,^{\pm}\!\mbox{\boldmath$\pi$}}

\newcommand{\pmbSig}{\,^{\pm}\!\mbox{\boldmath$\Sigma$}}
\newcommand{\pmSig}{\,^{\pm}\!\Sigma}
\newcommand{\pmbv}{\,^{\pm}\!\mbox{\boldmath$v$}}
\newcommand{\pmbw}{\,^{\pm}\!\mbox{\boldmath$w$}}
\newcommand{\pbv}{\,^{+}\!\mbox{\boldmath$v$}}
\newcommand{\mbv}{\,^{-}\!\mbox{\boldmath$v$}}
\newcommand{\pmv}{\,^{\pm}\!v}
\newcommand{\pmw}{\,^{\pm}\!w}
\newcommand{\bphi}{\mbox{\boldmath$\phi$}}

\newcommand{\sbl}{\mbox{\scriptsize\boldmath$l$}}
\newcommand{\br}{\mbox{\boldmath$r$}}
\newcommand{\sdef}{\stackrel{\rm def}{=}}
\newcommand{\ve}{\varepsilon}
\newcommand{\R}{\bar{R}}
\renewcommand{\t}{\textstyle}
\newcommand{\pmbtau}{\,^{\pm}\!\mbox{\boldmath$\tau$}}

\newcommand{\pbpi}{\,^+\!\mbox{\boldmath$\pi$}}
\newcommand{\mbpi}{\,^-\!\mbox{\boldmath$\pi$}}

\begin{document}

\title{Feynman path integral in area tensor Regge calculus and correspondence
principle}
\author{V.M. Khatsymovsky \\
 {\em Budker Institute of Nuclear Physics} \\ {\em
 Novosibirsk,
 630090,
 Russia}
\\ {\em E-mail address: khatsym@inp.nsk.su}}
\date{}
\maketitle
\begin{abstract}
The quantum measure in area tensor Regge calculus can be constructed in such the way
that it reduces to the Feynman path integral describing canonical quantisation if the
continuous limit along any of the coordinates is taken. This construction does not
necessarily mean that Lorentzian (Euclidean) measure satisfies correspondence
principle, that is, takes the form proportional to $e^{iS}$ ($e^{-S}$) where $S$ is
the action. Requirement to fit this principle means some restriction on the action,
or, in the context of representation of the Regge action in terms of independent
rotation matrices (connections), restriction on such representation. We show that the
representation based on separate treatment of the selfdual and antiselfdual rotations
allows to modify the derivation and give sense to the conditionally convergent
integrals to implement both the canonical quantisation and correspondence principles.
If configurations are considered such that the measure is factorisable into the
product of independent measures on the separate areas (thus far it was just the case
in our analysis), the considered modification of the measure does not effect the
vacuum expectation values.
\end{abstract}
\newpage
In our previous work \cite{Kha1} we have constructed the quantum measure in area
tensor Regge calculus with the following property. Whatever coordinate $t$ is chosen
along which the continu\-ous limit is taken, the resulting (properly defined)
continuous limit of the quantum measure is the Feynman path integral corresponding to
the canonical quantisation of the resulting system with continuous time $t$. The
possibility for such measure to exist is specific for the area tensor Regge calculus.
The latter reminds in this respect the 3-dimensional Regge calculus for which the same
problem has been solved in our earlier work \cite{Kha2}.

In three dimensions the constructed completely discrete measure takes the form
\begin{equation}                                                                    
\label{d-mu-3}%
\d\mu_3 = \exp\left (i\sum_{\sigma^1}\bl_{\sigma^1}*R_{\sigma^1}(\Omega)\right
)\prod_{\sigma^1\not\in{\cal F}}\d^3\bl_{\sigma^1}\prod_{\sigma^2}{\cal
D}\Omega_{\sigma^2}, ~~~ \bl*R\stackrel{\rm def}{=}{1\over 2}l^aR^{bc}\epsilon_{abc}.
\end{equation}

\noindent Here the field variables are the vectors $\bl_{\sigma^1}$ on the edges
$\sigma^1$ (the 1-dimensional simplices $\sigma^1$) and SO(3) connection matrices
$\Omega_{\sigma^2}$ on the triangles $\sigma^2$ (the 2-simplices $\sigma^2$). The
curvature matrix $R_{\sigma^1}(\Omega)$ is the (ordered along the path enclosing the
edge $\sigma^1$) product of the matrices $\Omega^{\pm 1}_{\sigma^2}$ for the triangles
$\sigma^2$ containing $\sigma^1$. The ${\cal F}$ is some set of edges arranged in
non-intersecting and non-self-intersecting broken lines passing through each vertex of
the manifold. Considered is the Euclidean signature case, and the integral in
(\ref{d-mu-3}) is defined by means of rotation of the integration contours in the
complex plane, $\bl$ $\to$ $-i\bl$ (to simplify notations, we do not show in
(\ref{d-mu-3}), but imply that this rotation is made only for those $\bl_{\sigma^1}$
which are the integration dummy variables, not for $\sigma^1$ $\in$ ${\cal F}$). The
${\cal D}\Omega$ is the Haar measure on SO(3). Arising it in the canonical approach is
connected with the structure of the Lagrangian originating from the Regge action in
the continuous time limit and resulting in the set of constraints which turn out to
coincide with those proposed for the 3-dimensional gravity by Waelbroeck \cite{Wael}
from symmetry considerations.

In four dimensions the result for the Euclidean measure being applied to arbitrary
function on the set of area tensors $\pi$ and connection matrices $\Omega$ reads
\begin{eqnarray}                                                                    
\label{VEV2}%
<\Psi (\{\pi\},\{\Omega\})> & = & \int{\Psi (-i\{\pi\}, \{\Omega\})\exp{\left (-\!
\sum_{\stackrel{t-{\rm like}}{\sigma^2}}{\tau _{\sigma^2}\circ
R_{\sigma^2}(\Omega)}\right )}}\nonumber\\
 & & \hspace{-20mm} \exp{\left (i
\!\sum_{\stackrel{\stackrel{\rm not}{t-{\rm like}}}{\sigma^2}} {\pi_{\sigma^2}\circ
R_{\sigma^2}(\Omega)}\right )}\prod_{\stackrel{\stackrel{\rm
 not}{t-{\rm like}}}{\sigma^2}}{\rm d}^6
\pi_{\sigma^2}\prod_{\sigma^3}{{\cal D}\Omega_{\sigma^3}} \nonumber\\ & \equiv &
\int{\Psi (-i\{\pi\},\{\Omega\}){\rm d} \mu_{\rm area}(-i\{\pi\},\{\Omega\})}, ~~~
\pi\circ R \stackrel{\rm def}{=}{1\over 2}\pi_{ab}R^{ab}.
\end{eqnarray}

\noindent We define tensors on the triangles $v^{ab}_{\sigma^2}$ (analogs of the
bivectors $\epsilon^{ab}{}_{cd}l^c_1l^d_2/2$ in ordinary Regge calculus formed by the
pairs of vectors $l^a_1$, $l^a_2$), now independent variables, and SO(4) connections
on the tetrahedra $\Omega_{\sigma^3}$. Here analog of the set ${\cal F}$ in
(\ref{d-mu-3}) consisting of the triangles $\sigma^2$ integration over tensors of
which is absent is specified. For that the regular method of constructing the
4-dimensional Regge manifold from the 3-dimensional Regge manifolds used in
\cite{Kha1} (analogous to that proposed in \cite{MisThoWhe}) is considered, these
3-dimensional manifolds usually being referred to as {\it the leaves of the
foliation}. A possible choice for the triangles of ${\cal F}$ are the $t$-like ones
where $t$ labels the sections, that is, those triangles one of edges of which is
located along the line of the coordinate $t$. Besides that, there are also the leaf
triangles completely contained in the 3-dimensional leaves and diagonal ones which are
neither leaf nor $t$-like triangles. Correspondingly, the tensors $v_{\sigma^2}$ are
divided into those of the $t$-like triangles $\tau_{\sigma^2}$ and of the leaf and
diagonal ones, $\pi_{\sigma^2}$. The integration is to be performed over only the
latter tensors, and the rotation of the integration contours is made just for them,
$\pi_{\sigma^2}$ $\to$ $-i\pi_{\sigma^2}$.

Another form of (\ref{VEV2}) convenient for calculation follows upon formal splitting
matrices into selfdual and antiselfdual parts,
\begin{eqnarray}                                                                    
\label{d-mu+-}%
{\rm d} \mu_{\rm area} & = & {\rm d} \,^+\!\mu_{\rm area}{\rm d} \,^-\!\mu_{\rm area},
\nonumber\\\lefteqn{{\rm d} \,^{\pm}\!\mu_{\rm area}(-i\{\pi\},\{\Omega\}) =
\exp{\left (-\! \sum_{\stackrel{t-{\rm like}}{\sigma^2}}{\pmtau _{\sigma^2}\circ
R_{\sigma^2}(\,^{\pm}\!\Omega)}\right )}}\nonumber\\
 & & \hspace{-15mm} \exp{\left (i
\!\sum_{\stackrel{\stackrel{\rm not}{t-{\rm like}}}{\sigma^2}} {\pmpi_{\sigma^2}\circ
R_{\sigma^2}(\,^{\pm}\!\Omega)}\right )}\prod_{\stackrel{\stackrel{\rm
 not}{t-{\rm like}}}{\sigma^2}}{\rm d}^3
\pmbpi_{\sigma^2}\prod_{\sigma^3}{{\cal D}\,^{\pm}\!\Omega_{\sigma^3}}.
\end{eqnarray}

\noindent The group property SO(4) = SU(2) $\!\times\!$ SU(2) is used by decomposing
generator $w$ of the connection $\Omega$ = $\exp w$ as well as any antisymmetric
matrix into self- and antiselfdual parts,
\begin{equation}                                                                    
w = \,^+\!w + \,^-\!w, ~~~ {1\over 2}\epsilon^{ab}{}_{cd}\,^\pm\!w^{cd} = \pm
\,^\pm\!w^{ab}.
\end{equation}

\noindent Correspondingly, $\Omega$ as well as $R$ are decomposed multiplicatively,
\begin{equation}                                                                    
\label{Omega-dual}%
\Omega = \,^+\!\Omega \,^-\!\Omega, ~~~ \,^\pm\!\Omega = \exp \,^\pm\!w.
\end{equation}

\noindent The basis of (anti-)selfdual matrices $\pmSig^k_{ab}$ ($k$ = 1, 2, 3) is
introduced such that $i\pmSig^k_{ab}$ satisfy Pauli matrices algebra. Thereby
(anti-)selfdual parts of a tensor $v_{ab}$, the $\pmv_{ab}$, are parameterised by the
3-vector components,
\begin{equation}                                                                    
\pmv_{ab} = {1\over 2}\pmv_k\pmSig^k_{ab} = {1\over 2}\pmbv\pmbSig_{ab}
\end{equation}

\noindent (so that
\begin{equation}                                                                    
|v|^2 \stackrel{\rm def}{=} v\circ v = {1\over 2}|\pbv|^2 + {1\over 2}|\mbv|^2
\end{equation}

\noindent ) as well as generators of the connections and curvatures,
\begin{equation}                                                                    
\,^{\pm}\!\Omega = \exp (\pmw_k\pmSig^k), ~~~ \,^{\pm}\!R = \exp
(\,^{\pm}\!\phi_k\pmSig^k).
\end{equation}

\noindent The latter in the given notations correspond to the SU(2) rotations by the
angles $|\pmbw|$, $|\,^{\pm}\!\bphi|$. At the same time, these matrices can be
considered in the adjoint representation, that is, as those acting on the vectors
$\pmv_k$ as SO(3) rotations by twice as much angles, for example,
\begin{equation}                                                                    
\exp (\pmw_k\pmSig^k)\pmv_l\pmSig^l\exp (-\pmw_m\pmSig^m) = \left ({\cal
O}(2\pmbw)\pmbv\right )_k\pmSig^k
\end{equation}

\noindent is rotation by the angle $2|\pmbw|$ around the axis $\pmbw$.

The presented formulas for the discrete measure contain the exponential factor
different from the Euclidean expression $e^{-S}$ (corresponding to the Lorentzian
$e^{iS}$) where $S$ is the action which in terms of $\Omega$ takes the form
\cite{Kha3}
\begin{equation}                                                                   
\label{S-l-Omega}%
S(\bl,\Omega)=\sum_{\sigma^1}{l_{\sigma^1}\arcsin{{\bl_{\sigma^1}*R_{\sigma^1}(\Omega)
\over l_{\sigma^1}}}}
\end{equation}

\noindent for three dimensions and
\begin{equation}                                                                   
\label{S-RegCon}%
S(v,\Omega) = \sum_{\sigma^2}{|v_{\sigma^2}|\arcsin{v_{\sigma^2}\circ
R_{\sigma^2}(\Omega)\over |v_{\sigma^2}|}}
\end{equation}

\noindent in the 4-dimensional case. The matter is that our derivation of the measure
in area tensor Regge calculus or in ordinary 3-dimensional Regge calculus includes at
some stage representation of the $\delta$-function as Fourier transform of unity. This
representation allows to raise the constraints from $\delta$-functions to exponent in
the Feynman path integral constructed in the canonical quantisation formalism and to
include them into the Lagrangian with the help of the Lagrange multipliers over which
integrations in the measure arise. Then the quantum measure in the completely discrete
Regge manifold can be found which reduces to the canonical one when passing to the
continuous limit along any of the coordinates, this coordinate beginning to play the
role of time in the canonical formalism.

In three dimensions representation of the following type was used,
\begin{equation}                                                                   
\int{e^{\textstyle i\bl*R}{\d^3\bl\over (2\pi)^3}} = \delta^3\left ({R-\R\over
2}\right ).
\end{equation}

\noindent Here $\bl$ in the LHS play the role of the Lagrange multipliers in the
Lagrangian. Since Regge action is the sum of the terms like $l\arcsin (R*\bl/l)$, not
$R*\bl$ to which it reduces upon using the equations of motion (that is, on-shell),
consider integral of the type
\begin{equation}                                                                   
\label{g-represent-delta}%
\int{e^{\textstyle ilg(R*\bl/l)}{\d^3\bl\over (2\pi)^3}}, ~~~ g(x) = -g(-x), ~~~
g^{\prime}(0) = 1,
\end{equation}

\noindent where the function $g(x)$ is analytical in the neighbourhood of zero. Let us
pass to spherical coordinates so that $\d^3\bl$ = $l^2\d l$ $\!\d o_{\sbl}$, divide
integral over the angles $\d o_{\sbl}$ into those ones over the upper and over the
lower hemispheres and represent the latter as the integral over the upper hemisphere
and over negative $l$,
\begin{eqnarray}                                                                   
\label{int-g-to-delta}%
\lefteqn{\int{e^{\t ilg(r\cos\theta) - \ve l^2}l^2\d l\sin\theta\d\theta\d\varphi} =
<\cos\theta = z>} \nonumber\\ & = & 2\pi\int\limits^1_0\d z\int\limits^{\infty}_0e^{\t
ilg(rz) - \ve l^2}l^2\d l + 2\pi\int\limits^0_{-1}\d z\int\limits^{\infty}_0e^{\t
ilg(rz) - \ve l^2}l^2\d l \nonumber\\ & = & 2\pi\int\limits^1_0\d
z\int\limits^{+\infty}_{-\infty}e^{\t ilg(rz) - \ve l^2}l^2\d l =
-(2\pi)^2\int\limits^1_0\delta^{\prime\prime}_{\ve}\left (g(rz)\right )\d z
\nonumber\\ & = & -{1\over
2}(2\pi)^2\int\limits^1_{-1}\delta^{\prime\prime}_{\ve}\left (g(rz)\right )\d z ~~~
\stackrel{\ve\to 0}{\longrightarrow} ~~~ -{1\over
2}(2\pi)^2\int\limits^1_{-1}\delta^{\prime\prime}\left (g(rz)\right )\d z,
\end{eqnarray}

\noindent where $r$ = $|\br|$, $r^a$ $\sdef$ $\epsilon^a{}_{bc}R^{bc}/2$. Here
\begin{equation}                                                                   
\delta\left (g(rz)\right ) = {\delta (rz)\over g^{\prime}(rz)} = \delta (rz),
\end{equation}

\noindent and a parameter $\ve$ $>$ 0 is introduced for temporary regularization of
the $\delta$-function to carefully take into account the integration limit $z$ = 0
(the edge of the hemispheres) of the intermediate integrations which belongs to the
support of the $\delta$-function. Thus, for certain way of calculation the
conditionally convergent integral (\ref{g-represent-delta}) does not depend on the
details of behaviour of the function $g(x)$ with the specified properties and
therefore is equal to such the integral also at $g(x)$ = $x$, that is, it yields an
alternative representation of the $\delta$-function which can be used at $g(x)$ =
$\arcsin x$ in order to get Regge action in the connection representation in the
exponential. As a result, the measure (\ref{d-mu-3}) gets modified,
\begin{equation}                                                                   
\label{d-mu-3-arcsin}%
\d\tilde{\mu}_3 = \exp\left
(i\sum_{\sigma^1}l_{\sigma^1}\arcsin{\bl_{\sigma^1}*R_{\sigma^1} (\Omega)\over
l_{\sigma^1}}\right )\prod_{\sigma^1\not\in{\cal
F}}\d^3\bl_{\sigma^1}\prod_{\sigma^2}{\cal D}\Omega_{\sigma^2}.
\end{equation}

It is easy to see that independence on $g(x)$ at such the way of calculation follows
also upon introducing a product of the vector components $l^a$ under the integral
sign, that is, for
\begin{equation}                                                                   
\label{g-represent-d-delta}%
\int{e^{\textstyle ilg(R*\bl/l)}l^{a_1}l^{a_2}\ldots l^{a_n}{\d^3\bl\over (2\pi)^3}},
~~~ g(x) = -g(-x), ~~~ g^{\prime}(0) = 1
\end{equation}

\noindent thus giving derivatives of the $\delta$-function. This can be used for
passing in the backward direction, from the measure (\ref{d-mu-3-arcsin}) to the
length expectation values. In particular, neglecting contribution of the edges of
${\cal F}$ (or $t$-like ones) in (\ref{d-mu-3-arcsin}) we have factorisation of the
measure and, according to the above said, we get the same result for the expectation
values \cite{Kha2} obtained using the measure (\ref{d-mu-3}). On the other hand, since
there is no integration over the edges of ${\cal F}$ in (\ref{d-mu-3-arcsin}), just
taking into account contribution of them would result in the difference between
consequences of the measures (\ref{d-mu-3-arcsin}) and (\ref{d-mu-3}).

Let us proceed with the 4-dimensional case. The integral
\begin{equation}                                                                   
\label{int-g-v}%
\int e^{\t i|v|g(R\circ v/|v|)}{\d^6v\over (2\pi)^6}, ~~~ g(x) = -g(-x), ~~~
g^{\prime}(0) = 1
\end{equation}

\noindent cannot be reduced to $\delta$-function since upon separating the integration
into those ones over $\d |v|$ and over $\d o_v$ (the measure on the 5-dimensional
sphere $v\circ v$ = 1 or body angle measure) the integration element takes the form
$|v|^5\d |v|$ $\!\d o_v$. Here $|v|^5$ analytically continues to the region $|v|$ $<$
0 as an odd function in contrast to $l^2$ in (\ref{int-g-to-delta}) which continues to
$l$ $<$ 0 in even manner. Therefore reducing (\ref{int-g-v}) to the $\delta$-function
in a way analogous to (\ref{int-g-to-delta}) is impossible. That is, modification of
the derivation of the measure (\ref{VEV2}) in such the way that the measure would
include the exponent of the action (\ref{S-RegCon}) turns out to be impossible.

However, it is possible to modify representation of the action. In the selfdual
representation \cite{Kha3} area tensors and rotation generators are replaced by their
selfdual parts. Excluding rotations by means of the equations of motion we get half of
the exact Regge action. Adding for symmetry the selfdual and antiselfdual
representations we get instead of (\ref{S-RegCon}) the expression
\begin{eqnarray}                                                                   
\label{(self)+(anti)SU(2)}%
S(\,^+\!v,\,^+\!\Omega) + S(\,^-\!v,\,^-\!\Omega) & = &
\sum_{\sigma^2}|\pbv_{\sigma^2}| \arcsin {\,^+\!v_{\sigma^2}\circ
R_{\sigma^2}(\,^+\!\Omega)\over |\pbv_{\sigma^2}|} \nonumber\\* & & +
\sum_{\sigma^2}|\mbv_{\sigma^2}| \arcsin {\,^-\!v_{\sigma^2}\circ
R_{\sigma^2}(\,^-\!\Omega)\over |\mbv_{\sigma^2}|}.
\end{eqnarray}

\noindent While $|\pbv_{\sigma^2}|$ = $|\mbv_{\sigma^2}|$ in the usual Regge calculus,
contributions of the selfdual and antiselfdual sectors are completely independent in
the area tensor Regge calculus. Note that the representa\-tion in terms of the full
SO(4) rotations (\ref{S-RegCon}) can be rewritten identically by means of splitting
the matrices into selfdual and antiselfdual parts as
\begin{equation}                                                                   
\label{(self+anti)}%
S(v,\Omega) = \sum_{\sigma^2}|v_{\sigma^2}|\arcsin{\,^+\!v_{\sigma^2}\circ
R_{\sigma^2}(\,^+\!\Omega) + \,^-\!v_{\sigma^2}\circ R_{\sigma^2}(\,^-\!\Omega)\over
|v_{\sigma^2}|}.
\end{equation}

\noindent For small curvature when we can replace $\arcsin x$ $\to$ $x$ the difference
between (\ref{(self)+(anti)SU(2)}) and (\ref{(self+anti)}) vanishes.

The matrices $\,^{\pm}\!\Omega$ and $R(\,^{\pm}\!\Omega)$ in the equation
(\ref{(self)+(anti)SU(2)}) mean SU(2) rotations. These matrices in the SO(3)
representation are the rotations by twice as much angles. Taking into account this
fact we come to yet another representation of the Regge action as in the 3-dimensional
case,
\begin{eqnarray}                                                                   
\label{(self)+(anti)SO(3)}%
{1\over 2}S(\pbv,\,^+\!\Omega) + {1\over 2}S(\mbv,\,^-\!\Omega) & = & {1\over 2}
\sum_{\sigma^2}|\pbv_{\sigma^2}| \arcsin {\pbv_{\sigma^2} *
R_{\sigma^2}(\,^+\!\Omega)\over |\pbv_{\sigma^2}|} \nonumber\\* & & + {1\over 2}
\sum_{\sigma^2}|\mbv_{\sigma^2}| \arcsin {\mbv_{\sigma^2} *
R_{\sigma^2}(\,^-\!\Omega)\over |\mbv_{\sigma^2}|}.
\end{eqnarray}

The actions (\ref{(self)+(anti)SU(2)}), (\ref{(self+anti)}) and
(\ref{(self)+(anti)SO(3)}) are {\it different}, but reducable to the same {\it exact}
Regge action upon excluding connections by means of the equations of motion. If
connection matrices tend to unity, these actions tend to the equivalent ones. In
particular, this takes place in the continuum limit when we ascribe to the different
values the same orders of magnitude as would be on the ordinary Regge calculus
manifold obtained from a fixed smooth Riemannian manifold by triangulating it with
typical lattice spacing tending to zero. As a result, the continuum analogs of
(\ref{(self)+(anti)SU(2)}), (\ref{(self+anti)}) and (\ref{(self)+(anti)SO(3)}) are
equivalent.

Using the representations (\ref{(self)+(anti)SU(2)}) or (\ref{(self)+(anti)SO(3)})
allows to modify derivation of the measure to satisfy the correspondence principle
since the 3-dimensional $\delta$-functions $\delta^3(\,^+\!R$ $-$ $\!^+\!\bar{R})$,
$\delta^3(\,^-\!R$ $-$ $\!^-\!\bar{R})$ and their derivatives can be represented by
the expression of the type of (\ref{g-represent-d-delta}) so that a certain way of
calculating make the integrals
\begin{equation}                                                                   
\label{v-g-deltaSU(2)}%
\int e^{\t i|\pmbv|g(\,^{\pm}\!R\circ\pmv/|\pmbv|)}
\pmv^{k_1}\pmv^{k_2}\ldots\pmv^{k_n} {\d^3\pmbv\over (2\pi)^3}, ~~~ g(x) = -g(-x), ~~~
g^{\prime}(0) = 1,
\end{equation}

\noindent or
\begin{equation}                                                                   
\label{v-g-deltaSO(3)}%
\int e^{\t i|\pmbv|g(\,^{\pm}\!R * \pmbv/|\pmbv|)/2}
\pmv^{k_1}\pmv^{k_2}\ldots\pmv^{k_n} {\d^3\pmbv\over (2\pi)^3}, ~~~ g(x) = -g(-x), ~~~
g^{\prime}(0) = 1,
\end{equation}

\noindent independent on the details of behaviour of the analytical function $g(x)$.
We can use this circumstance in two directions. First, to get the following form of
the quantum measure modifying (\ref{VEV2}) (or (\ref{d-mu+-})),
\begin{eqnarray}                                                                   
\label{d-mu+-SU(2)}%
{\rm d} \mu^{\rm SU(2)}_{\rm area} & = & {\rm d} \,^+\!\mu^{\rm SU(2)}_{\rm area}{\rm
d} \,^-\!\mu^{\rm SU(2)}_{\rm area}, \nonumber\\ {\rm d} \,^{\pm}\!\mu^{\rm
SU(2)}_{\rm area}(-i\{\pi\},\{\Omega\}) & = & \exp{\left (-\! \sum_{\stackrel{t-{\rm
like}}{\sigma^2}}{|\pmbtau_{\sigma^2}|\arcsin{\pmtau _{\sigma^2}\circ
R_{\sigma^2}(\,^{\pm}\!\Omega)\over |\pmbtau_{\sigma^2}|}}\right )}\nonumber\\
 & & \hspace{-40mm} \exp{\left (i
\!\sum_{\stackrel{\stackrel{\rm not}{t-{\rm like}}}{\sigma^2}}
{|\pmbpi_{\sigma^2}|\arcsin{\pmpi_{\sigma^2}\circ R_{\sigma^2}(\,^{\pm}\!\Omega)\over
|\pmbpi_{\sigma^2}|}}\right )}\prod_{\stackrel{\stackrel{\rm
 not}{t-{\rm like}}}{\sigma^2}}{\rm d}^3
\pmbpi_{\sigma^2}\prod_{\sigma^3}{{\cal D}\,^{\pm}\!\Omega_{\sigma^3}}
\end{eqnarray}

\noindent or
\begin{eqnarray}                                                                   
\label{d-mu+-SO(3)}%
{\rm d} \mu^{\rm SO(3)}_{\rm area} & = & {\rm d} \,^+\!\mu^{\rm SO(3)}_{\rm area}{\rm
d} \,^-\!\mu^{\rm SO(3)}_{\rm area}, \nonumber\\ {\rm d} \,^{\pm}\!\mu^{\rm
SO(3)}_{\rm area}(-i\{\pi\},\{\Omega\}) & = & \exp{\left (-{1\over 2}\!
\sum_{\stackrel{t-{\rm like}}{\sigma^2}}{|\pmbtau_{\sigma^2}|\arcsin{\pmbtau
_{\sigma^2} * R_{\sigma^2}(\,^{\pm}\!\Omega)\over |\pmbtau_{\sigma^2}|}}\right
)}\nonumber\\
 & & \hspace{-40mm} \exp{\left ({i\over 2}
\!\sum_{\stackrel{\stackrel{\rm not}{t-{\rm like}}}{\sigma^2}}
{|\pmbpi_{\sigma^2}|\arcsin{\pmbpi_{\sigma^2} * R_{\sigma^2}(\,^{\pm}\!\Omega)\over
|\pmbpi_{\sigma^2}|}}\right )}\prod_{\stackrel{\stackrel{\rm
 not}{t-{\rm like}}}{\sigma^2}}{\rm d}^3
\pmbpi_{\sigma^2}\prod_{\sigma^3}{{\cal D}\,^{\pm}\!\Omega_{\sigma^3}}.
\end{eqnarray}

\noindent Second, to estimate the vacuum expectation values provided by such the
measure.

Applying (\ref{d-mu+-SU(2)}) or (\ref{d-mu+-SO(3)}) to estimating the vacuum
expectations we use independence of the integral (\ref{v-g-deltaSU(2)}) or
(\ref{v-g-deltaSO(3)}) on the details of $g(x)$ which allows to reduce it to the
simple case $g(x)$ = $x$. In particular, if we restore the integral over SU(2)
rotations
\begin{equation}                                                                   
\int e^{\t i|\pmbpi|g(\,^{\pm}\!R\circ\pmpi/|\pmbpi|)}{\cal D}\,^{\pm}\!R
\end{equation}

\noindent as function of $(\pmbpi)^2$ from the moments of it, that is, the results of
integrating it with powers of $(\pmbpi )^2$ (integrating first over $\pmbpi$, then
over $\,^{\pm}\!R$), we get the same result as at $g(x)$ = $x$. If we analogously
define the integral over SO(3) rotations
\begin{equation}                                                                   
\int e^{\t i|\pmbpi|g(\,^{\pm}\!R * \pmbpi/|\pmbpi|)/2}{\cal D}\,^{\pm}\!R,
\end{equation}

\noindent the result is again the same as at $g(x)$ = $x$. Since thus far our analysis
has been done when the measure factorises over separate areas, that is, disregarding
contribution of the $t$-like triangles, modification of the measure to the case $g(x)$
$\neq$ $x$ does not change the results.

Namely, representation in terms of SU(2) rotations leads, analogously to
\cite{Kha1,Kha2}, to the following expectation of a function on a single area,
\begin{eqnarray}                                                                   
\label{Euclidean_measure}%
<f(\pi)> = \int{f (-i\pi ){\rm d}^6\pi\int{e^{\textstyle i\pi\circ R}{\cal
D}R}}&=&\int{f (\pi ) {\nu_2 (|\pbpi |)\over |\pbpi |^2}{\nu_2 (|\mbpi |)\over |\mbpi
|^2} {\d^3\pbpi\over 4\pi} {\d^3\mbpi\over 4\pi}}, \nonumber\\ \nu_2(l)={l\over\pi}
\int\limits_{0}^{\pi}{{\rm d}\varphi\over\sin^2{\!\varphi}}\,{e^{\textstyle
-l/\sin{\varphi}}}. & &
\end{eqnarray}

If, on the other hand, we issue from the representation in terms of SO(3) rotations,
(\ref{(self)+(anti)SO(3)}), expectation of the function of area follows from
(\ref{Euclidean_measure}) by replacing $\nu_2 (|\pmbpi|)$ $\to$ $\nu_3 (|\pmbpi|/2)$
and (for normalisation) $\d^3\pmbpi$ $\to$ $\d^3\pmbpi /2$ where $\nu_3$ arises in the
3-dimensional case \cite{Kha2} when deriving expectation of a function on an edge,
\begin{eqnarray}                                                                   
\label{<g>}%
<f(\bl)> = \int{{{\rm d}o_{\sbl} \over 4\pi}\int\limits_{0}^{\infty}{f(\bl)\nu_3
(l){\rm d}l}}, ~~~ \nu_3 (l)={2l \over \pi}\int\limits_{0}^{\pi}{\exp{\left( -{l \over
\sin{\varphi}} \right){\rm d}\varphi}}.
\end{eqnarray}

\noindent The recipe of averaging a function of an area given in our work \cite{Kha1}
and further used in \cite{Kha4} to reduce the measure to the physical surface just
corresponds to this representation (that is, to the measure (\ref{d-mu+-SO(3)}) or to
that one with the replacement $\arcsin x$ $\to$ $x$ made which, as considered above,
does not effect the result in the framework of the factorisation). The difference
between $\nu_3$ and $\nu_2$ is due to the difference between the elements of the Haar
measure for SO(3) and SU(2) rotations by the same angle $\bphi$,
\begin{equation}                                                                   
\label{3d-Haar}%
{\cal D}R={\sin^2(\phi /2) \over 4\pi^2\phi^2}{\rm d}^3\bphi
\parbox{2cm}{~~~ or} {\sin^2\phi \over 2\pi^2\phi^2}{\rm d}^3\bphi,
\end{equation}

\noindent respectively, the antisymmetric part of $R$ (which just appears in the
exponential in the measure) being proportional to $\sin \phi$ in both cases.

If taking into account contribution of the $t$-like triangles (the problem thus far
not solved explicitly in general case) the choices $g(x)$ = $x$ and $g(x)$ $\neq$ $x$
can lead to the different consequences since the scale of the $t$-like triangles is
fixed (is not a dummy integration variable). This really displays the difference
between the measures (\ref{VEV2}) (or (\ref{d-mu+-})) and (\ref{d-mu+-SU(2)}) and also
between the measure (\ref{d-mu+-SO(3)}) and that one following from it upon replacing
$\arcsin x$ $\to$ $x$.

Thus, it is possible to satisfy, in addition to the principle of canonical
quantisation and equivalence of coordinates, also the correspondence principle
(proportionality of the Lorentzian (Euclidean) measure to $e^{iS}$ ($e^{-S}$), $S$
being the action) when constructing quantum measure in area tensor Regge calculus or
in 3-dimensional Regge calculus. In four dimensions we should use representation of
the Regge action with decomposition into independent contributions of selfdual and
antiselfdual sectors of the theory for that. Possible are the two versions of such the
representation based on decomposing SO(4) connections as SU(2) $\!\times\!$ SU(2) or
SO(3) $\!\times\!$ SO(3). Modification of the quantum measure does not effect the
results of estimating vacuum expectation values of a function of a single area
obtained in suggestion $|\pmbtau|$ = 0 when factorisa\-tion into measures on the
separate areas is possible.

Recently it turns out to be possible to average a function on areas with the help of
the measures (\ref{d-mu+-}), (\ref{d-mu+-SU(2)}) and (\ref{d-mu+-SO(3)}) on some
simplest configurations of area tensor Regge calculus at $|\pmbtau|$ $\neq$ 0
\cite{Kha5}. All the measures turn out to have probabilistic interpretation (positive)
on the physical surface (where $|\pbv|$ = $|\mbv|$) on these configurations and have a
smooth limit at $|\pmbtau|$ $\to$ 0 coinciding with the known $|\pmbtau|$ = 0 results
(\ref{Euclidean_measure}), (\ref{<g>}), however, the type of the dependence on
$\pmbtau$ is qualitatively different for the different measures.

\bigskip

The present work was supported in part by the Russian Foundation for Basic Research
through Grant No. 03-02-17612.

\end{document}